\documentclass[twocolumn]{aastex631}
\AtBeginDocument{\expandafter\let\csname internallinenumbers\endcsname\relax\expandafter\let\csname endinternallinenumbers\endcsname\relax}

\usepackage{amsmath}
\usepackage{xcolor}
\newcommand{\rd}[1]{#1}

\newcommand{\kms}{km\,s$^{-1}$}
\newcommand{\um}{\,$\mu$m}

\shorttitle{H$_3^+$ in the Ring Nebula}
\shortauthors{Neufeld et al.}

\begin{document}

\title{H$_3^+$ forbidden rotational lines in a planetary nebula:\\ JWST/MIRI detection in the Ring Nebula and ionization by the central star}

\author[0000-0001-8341-1646]{David A. Neufeld}
\affiliation{William H. Miller III Department of Physics and Astronomy, Johns Hopkins University, 3400 North Charles Street, Baltimore, MD 21218, USA}
\email{neufeld@jhu.edu}

\author[0000-0003-2027-5020]{Paule Sonnentrucker}
\affiliation{European Space Agency (ESA) Office, Space Telescope Science Institute, 3700 San Martin Drive, Baltimore, MD 21218, USA}

\author[0000-0003-1665-5709]{Joel D. Green}
\affiliation{Space Telescope Science Institute, 3700 San Martin Drive, Baltimore, MD 21218, USA}

\begin{abstract}
We report the detection of the forbidden pure rotational spectrum of H$_3^+$ in the Ring Nebula (NGC 6720), to our knowledge the first
detection of H$_3^+$ in a planetary nebula. In JWST/MIRI spectra obtained toward the inner edge of the main ring, 0.1\,pc in
projection from the central star, the $(J,K) = (5,0)\to(4,3)$ line at 16.325\um\ is detected at $32\,\sigma$, along with nine further unblended
lines. The H$_3^+$ emission spatially follows the H$_2$ emission, whose rotational lines show warm (660\,K) and hot (1700\,K) components.
A statistical-equilibrium model at these temperatures reproduces the subthermal H$_3^+$ line pattern, with much of the emission arising in the hot H$_2$. The emission requires an H$_2$
ionization rate of $\sim 3\times10^{-12}$\,s$^{-1}$ in the hot H$_2$ for an electron fraction of $2\times10^{-4}$. The rate is $\sim 2\times10^4$ times the cosmic-ray ionization rate in diffuse clouds. The
extreme-ultraviolet and soft X-ray radiation (54--300\,eV) of the 135\,kK central star supplies this rate if the hot H$_2$ lies within
\rd{an H-nucleus} column density of $(4$--$8)\times10^{19}$\,cm$^{-2}$ of the ionization front of the dense globules. Static photoionization models\rd{,} however, place the H$_2$ in gas too cold to excite the H$_3^+$ lines. The warm molecular gas and its H$_3^+$ thus suggest
dynamical effects, most plausibly the advection of H$_2$ through a photoevaporation front into gas heated by the
star. H$_3^+$ provides a new probe of the
photoionization of molecular gas by \rd{very} hot stars.
\end{abstract}


\section{Introduction} \label{sec:intro}

H$_3^+$ is formed when H$_2$ is ionized and the resulting H$_2^+$ reacts with H$_2$. In molecular clouds the ionization is provided by
cosmic rays, and H$_3^+$ absorption in its near-infrared vibrational bands has become a primary measure of the cosmic-ray ionization rate
\citep{Geballe1996,McCall2003,Oka2006,Indriolo2012,Miller2020}. \rd{H$_3^+$ vibrational lines have also been detected
in emission with JWST/NIRSpec in luminous infrared galaxies \citep{PereiraSantaella2024}, and a likely detection has also been obtained
toward the irradiated protoplanetary disk d203-506 in Orion \citep{Schroetter2025,Goicoechea2025}.}

Because H$_3^+$ has no permanent dipole moment, its pure rotational
transitions are ``forbidden''; they arise only through centrifugal distortion \citep{Watson1971,Pan1986}. \citet{Draine1990} predicted that the strongest of
them, the $(J,K) = (5,0)\to(4,3)$ line at 16.33\um, would be emitted by X-ray-irradiated molecular gas in the starburst galaxy
NGC 6240. With JWST/MIRI we recently detected this forbidden
rotational spectrum for the first time, in the shocked molecular gas of HH 168 \citep{Neufeld2026a}.

Planetary nebulae (PNe) offer a very different environment for H$_3^+$. Evolved PNe such as the Ring Nebula (NGC 6720) contain thousands
of dense molecular globules, bright in H$_2$ emission, that are exposed to the radiation of a hot central star \citep{Wesson2024}. The
star's extreme-ultraviolet (EUV) and soft X-ray photons penetrate the neutral gas and can ionize H$_2$ directly. The ionized gas of the Ring Nebula emits the fine-structure lines of a wide range of ions, including [Ar\,{\sc ii}], [Ne\,{\sc ii}], [S\,{\sc iii}], [Ar\,{\sc iii}], [S\,{\sc iv}], [Cl\,{\sc iv}], [Ne\,{\sc iii}] and [O\,{\sc iv}] \citep{Wesson2024,Clark2025}, with appearance potentials of 16--55\,eV.

Several molecular ions
have been detected in PNe: HeH$^+$ in NGC 7027 \citep{Gusten2019,Neufeld2020}, CH$^+$ in IC 4997 \citep{Rao2020}, CH$_3^+$ in NGC 6302
\citep{Bhatt2025}, N$_2$H$^+$ in NGC 7027 \citep{Zhang2008}, and OH$^+$, observed with Herschel in the Ring Nebula itself and in other
PNe with central stars hotter than $10^5$\,K \citep{Aleman2014,Etxaluze2014}. The N$_2$H$^+$ in NGC 7027 has been taken to imply H$_3^+$
formation by soft X-rays \citep{Zhang2008}, and the OH$^+$ may be driven by the
stellar high-energy photons \citep{Aleman2014}. H$_3^+$ itself, however, has not been detected in any PN. Photoionization models of PNe without dense clumps, in
which the H$_2$ lies in the ionized nebula and the transition zone at its edge, predict negligible H$_3^+$ abundances, $\lesssim 3\times10^{-13}$
\citep{Aleman2004,vanHoof2010}.

Here we report the detection of the H$_3^+$ rotational spectrum in archival JWST/MIRI observations of the Ring Nebula, and show that
photoionization by the central star can supply the required ionization rate.

\section{Observations and data reduction} \label{sec:obs}

\begin{figure*}[!htbp]
\centering\includegraphics[width=0.8\textwidth]{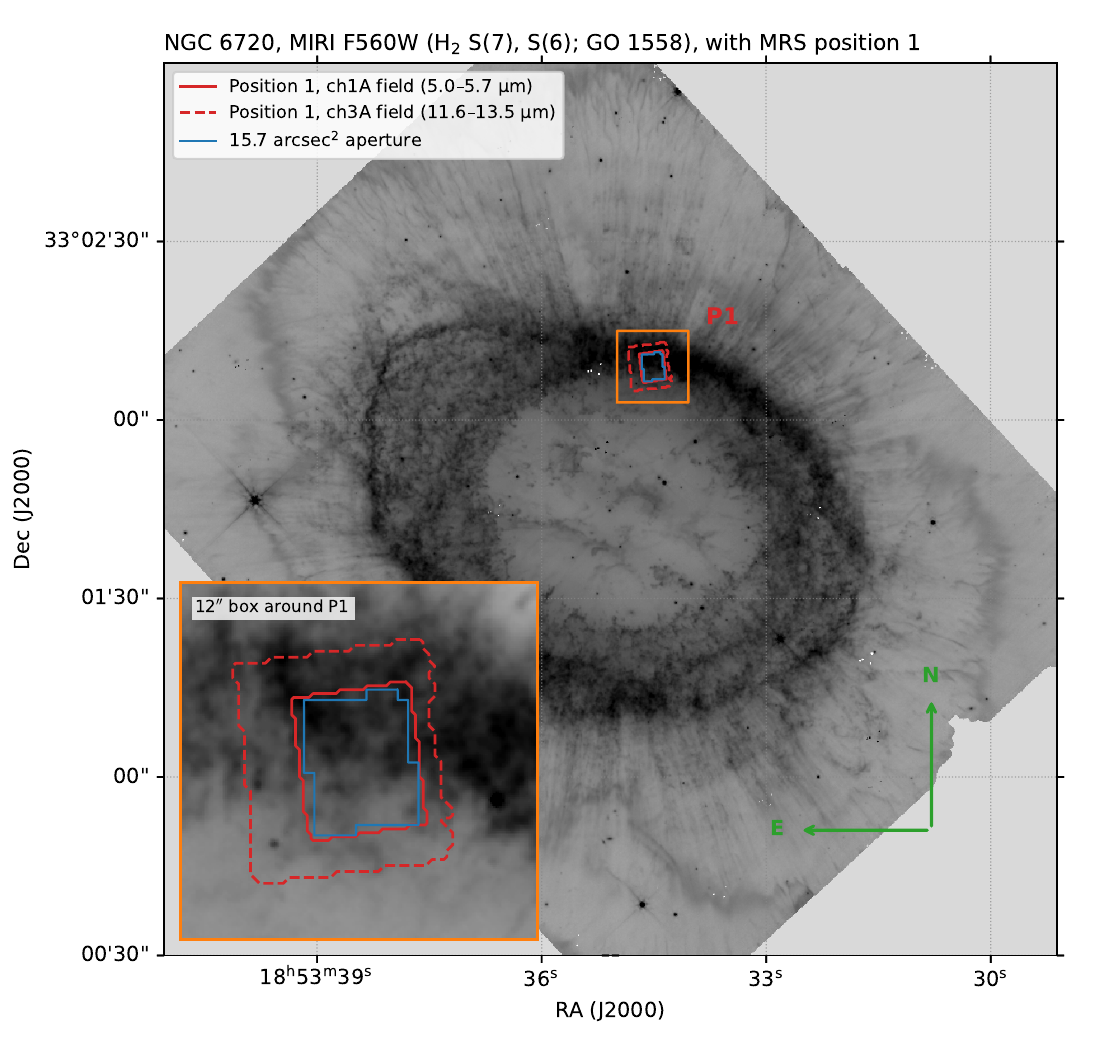}
\caption{The Ring Nebula in the MIRI F560W image from program GO 1558, which includes the H$_2$ S(7) and S(6) lines, with the MRS
position 1 fields in channel 1A (solid red) and channel 3A (dashed red) and the 15.7\,arcsec$^2$ aperture common to all sub-bands
(blue). The inset shows the $12\arcsec \times 12\arcsec$ region around the field, outlined in orange in the main panel. North is up and east to the left.\label{fig:field}}
\end{figure*}

The Ring Nebula was observed with the MIRI Medium-Resolution Spectrometer \citep[MRS;][]{Argyriou2023} on 2022 July 4 as part of JWST
program GO 1558 (PI M.~Barlow; \citealt{Wesson2024,Clark2025,Sahai2025}). We use MRS position 1 (observation 2), centered at
$\alpha = 18^{\rm h}53^{\rm m}34\fs51$, $\delta = +33\arcdeg02\arcmin09\farcs1$ (J2000), on the inner edge of the main ring, 25.0\arcsec\
(0.096\,pc in projection, for a distance of 790\,pc; \citealt{Wesson2024}) from the central star (Figure~\ref{fig:field}). All 12
sub-bands were observed with 921\,s each. We used the stage-3 cubes from the Mikulski Archive for Space Telescopes (MAST; pipeline version 2.0.1, context jwst\_1535.pmap).

The analysis follows \citet{Neufeld2026a}. The cubes were convolved to a common resolution of 0.87\arcsec, the MRS resolution at
23\um\ as in that paper, and regridded to a common 0.35\arcsec\ grid. The H$_2$ lines, which are bright, were fitted with Gaussians of free center and width. The H$_3^+$ lines were fitted with
Gaussians of fixed center and fixed width on a quadratic baseline, with cataloged
lines of other species within the fit window included as additional components. The fixed width in each MIRI sub-band is that
measured from the brightest unblended lines, which are unresolved (FWHM 85--100\,\kms\ in channels 1--3 and 130--140\,\kms\ in
channel 4; \citealt{Neufeld2026a}). The fixed center is the velocity of the 16.325\um\ line, from a single-Gaussian fit, with free center and width,
to that line in the spectrum summed over the 15.7\,arcsec$^2$ aperture defined below. That free fit only sets the center: all H$_3^+$
fluxes and maps, including those of the 16.325\um\ line, are from the fixed-center, fixed-width fits. The fits
were made in each spaxel for the maps and to the summed spectrum of the aperture for the fluxes. Repeating the fits at
line-free wavelengths showed that the formal uncertainties are underestimated by factors of 1.2--1.4 in channel 2 and 1.16 in
sub-band 3A; they were enlarged accordingly.

The cubes are not background-subtracted, but each line is measured above a local baseline. We extracted fluxes in the 15.7\,arcsec$^2$ aperture (128 pixels) covered by all 12 sub-bands (Figure~\ref{fig:field}), so that
the H$_2$ S(1)--S(8) lines and all H$_3^+$ lines are measured over the same area. The H$_2$ and H$_3^+$ fluxes are listed in Table~\ref{tab:fluxes}
(Appendix~\ref{app:tables}). Fluxes were corrected for extinction with the KP5 curve \citep{Pontoppidan2024,Chapman2009}, with $A_V$
determined from the H$_2$ lines (Section~\ref{sec:h2}). Earlier mid-infrared studies of the H$_2$ and PAH emission of the Ring Nebula applied no extinction
correction \citep{Cox2016,Clark2025}; here the correction is small, a factor of 1.14 at 16.325\um\ for $A_V = 1.7 \pm 0.9$\,mag.

\section{Results} \label{sec:results}

\subsection{\texorpdfstring{H$_3^+$}{H3+} lines} \label{sec:lines}

\begin{figure*}[!htbp]
\plotone{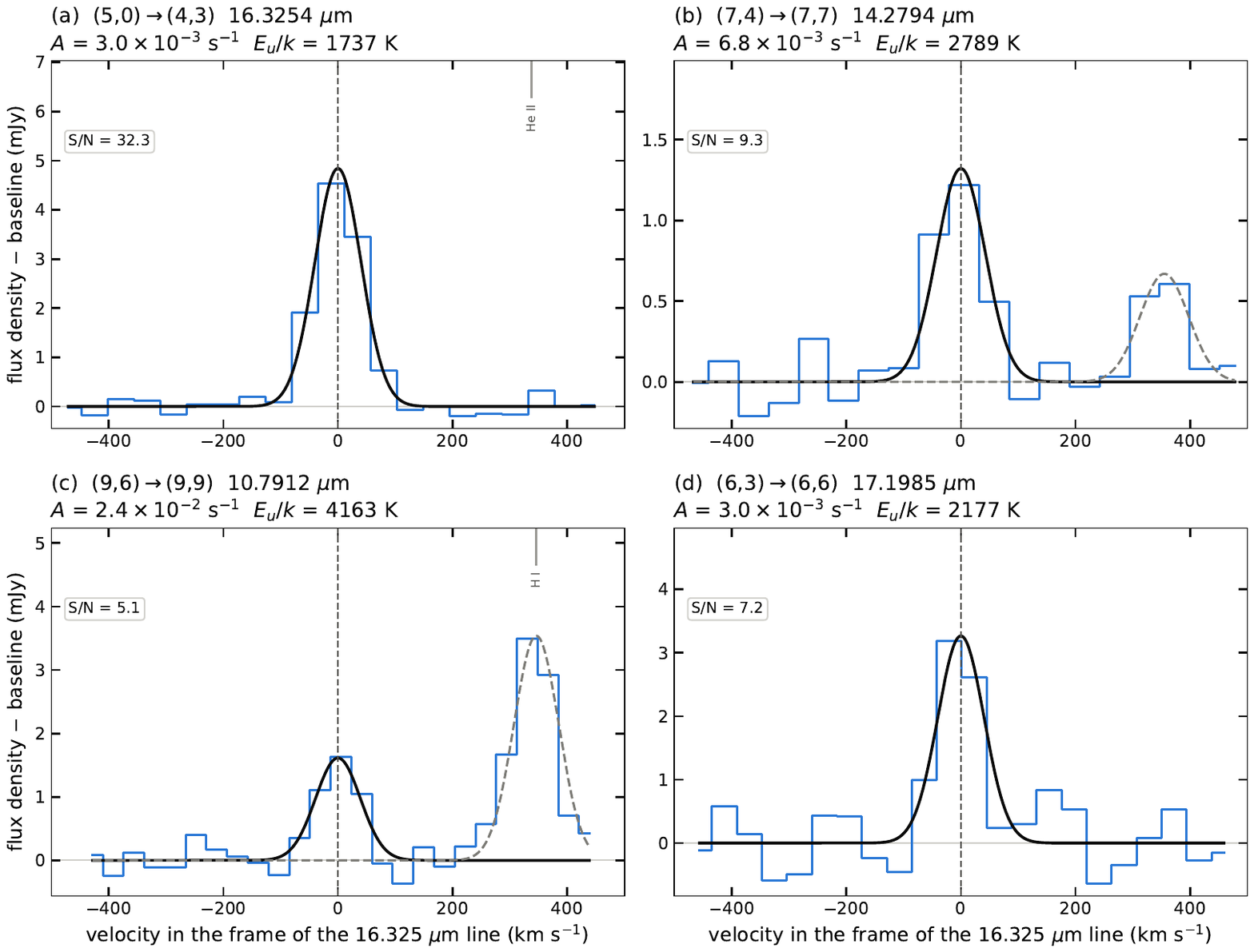}
\caption{The four unblended H$_3^+$ lines detected with the highest S/N in the 15.7\,arcsec$^2$ aperture, against velocity in the frame defined by the fitted center of the 16.325\um\ line, after
subtraction of a quadratic baseline. Black: Gaussian fits of fixed center and fixed width (Section~\ref{sec:obs}); dashed: neighboring lines fitted
simultaneously (an unidentified line in panel b, H\,{\sc i} 16--9 in panel c); the H$_3^+$ $(12,5)\to(11,4)$ transition at 14.296\um, with $E_u/k = 7945$\,K, is predicted to be $\sim 10^4$ times weaker than the 16.325\um\ line. Histograms: observed spectra. Gray ticks mark cataloged lines of other species.\label{fig:spectra}}
\end{figure*}

The H$_3^+$ $(5,0)\to(4,3)$ line at 16.325\um, the strongest H$_3^+$ line in HH 168, is detected at $32\,\sigma$
(Figure~\ref{fig:spectra}). We searched in the Ring Nebula for the 26 H$_3^+$ lines detected in HH 168 (Table 1 of \citealt{Neufeld2026a}), all of which are covered by the Ring Nebula observations. Four are blended here with [S\,{\sc iv}], H\,{\sc i} 18--11 or each other. Of the remaining 22 unblended lines (Table~\ref{tab:fluxes}), ten are detected, four of them above $5\,\sigma$ (16.325, 14.279, 17.198 and 10.791\um), and twelve are upper limits (Figure~\ref{fig:excitation}).

The centroid of the 16.325\um\ line, from the free fit of Section~\ref{sec:obs}, lies $+2.2 \pm 1.4$\,\kms\
from that of H$_2$ S(7) and $+7.1 \pm 1.4$\,\kms\ from that of H$_2$ S(1) in the
same sub-band (statistical uncertainties). The H$_3^+$ line is thus at the H$_2$ velocity to within $\sim 7$\,\kms; its width
(100\,\kms\ FWHM), like that of H$_2$ S(1) (97\,\kms), is close to the fixed width adopted for this sub-band (95\,\kms), so
the line is unresolved.

The identification follows from the same arguments as for HH 168 \citep{Neufeld2026a}: the lines fall at
the MiZATeP wavelengths \rd{\citep{Mizus2017,Bowesman2023}} of the transitions detected there, and no other plausible carrier is known at
16.325\um\ in this field. The 16.325\um\ line is present in every dither position and absent from the dedicated background field.

\begin{figure*}[!htbp]
\centering\includegraphics[width=0.78\textwidth]{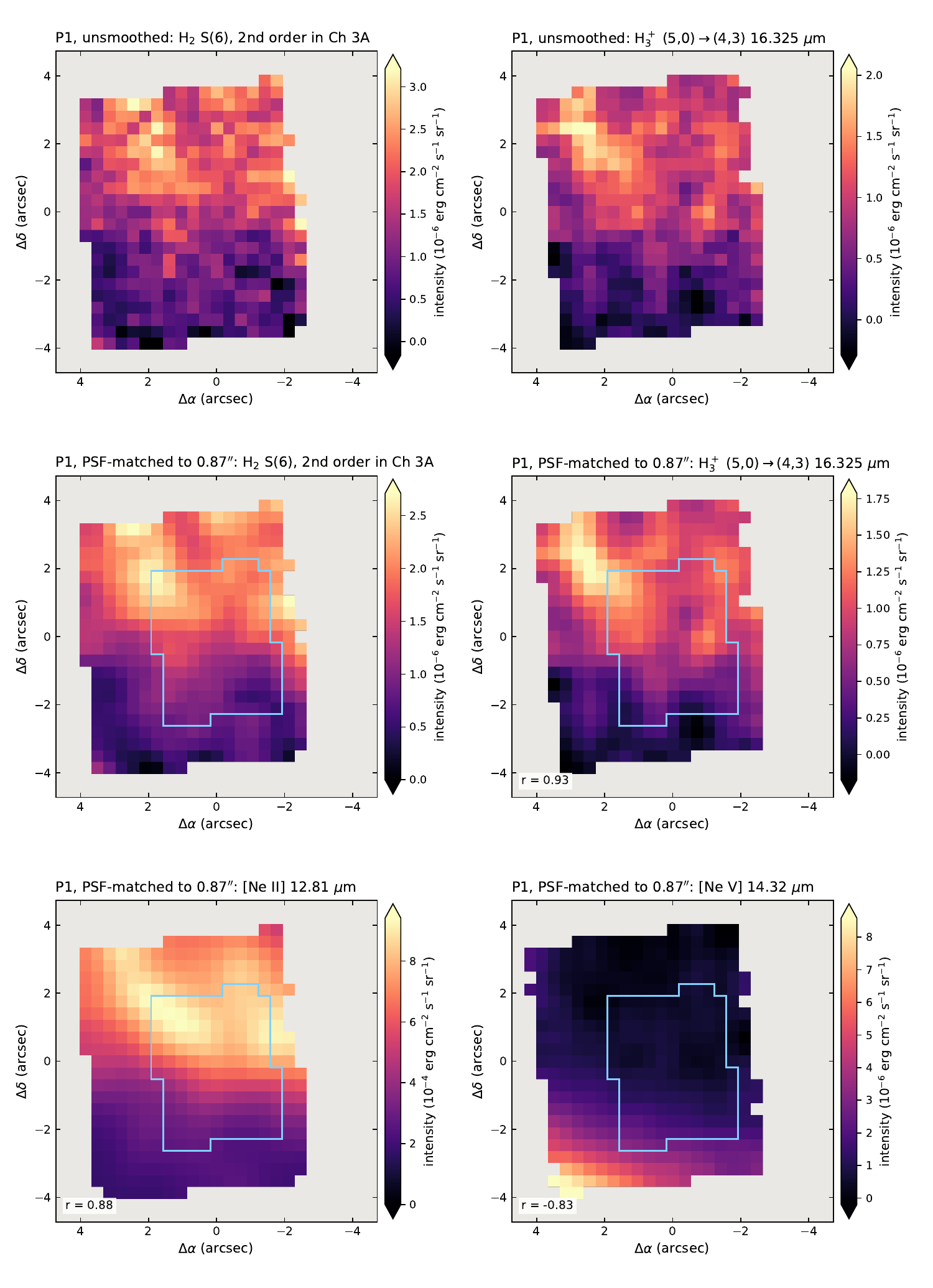}
\caption{Maps of H$_2$ S(6) observed in second order in channel 3A (left) and of the H$_3^+$ 16.325\um\ line (right), unsmoothed (top) and
convolved to 0.87\arcsec\ (middle), and of the ionic lines [Ne\,{\sc ii}] 12.81\um\ and [Ne\,{\sc v}] 14.32\um, convolved to
0.87\arcsec\ (bottom). The outline (middle and bottom) marks the 15.7\,arcsec$^2$ aperture. $r$ is the pixel-by-pixel correlation
coefficient between the convolved H$_3^+$ map and the first-order H$_2$ S(6) map from channel 1 (not shown), or the ionic line
shown (bottom), over the pixels where the
16.325\um\ line is detected above $3\,\sigma$. Offsets are from the field center.\label{fig:maps}}
\end{figure*}

The H$_3^+$ emission follows the H$_2$ (Figure~\ref{fig:maps}). We compared the 16.325\um\ map with H$_2$ S(6), observed both in first
order in channel 1 and in second order in channel 3A \citep{Neufeld2026b}; the latter has the same field as the H$_3^+$ line. Pixel by
pixel, over the pixels where the 16.325\um\ line is detected above $3\,\sigma$, its intensity correlates with the channel 1 S(6)
intensity with a correlation coefficient $r = 0.93$. The S(6) emission comes mostly from the hot H$_2$ component; the correlation with
S(1), which traces the warm component, is weaker ($r = 0.76$ over the same pixels). Over these pixels the H$_3^+$
emission is anticorrelated with the high-ionization lines [Ne\,{\sc v}] 14.32\um\ and [S\,{\sc iv}] 10.51\um\
($r = -0.83$ and $-0.82$), and correlated with [Ne\,{\sc ii}] 12.81\um\ ($r = 0.88$) and with the other low-ionization lines [Ar\,{\sc ii}] 6.985 and [Fe\,{\sc ii}] 5.340\um\ ($r = 0.86$ and 0.76), as is the H$_2$ S(6) emission, whereas the bright intermediate-ionization lines [Ne\,{\sc iii}] 15.56 and [S\,{\sc iii}] 18.71\um\ are uncorrelated or anticorrelated with it ($r = +0.06$ and $-0.55$);
these correlations largely reflect the ionization gradient across the field.

\subsection{\texorpdfstring{H$_2$ and H$_3^+$}{H2 and H3+} excitation} \label{sec:h2}

\begin{figure*}[!htbp]
\centering\includegraphics[width=0.62\textwidth]{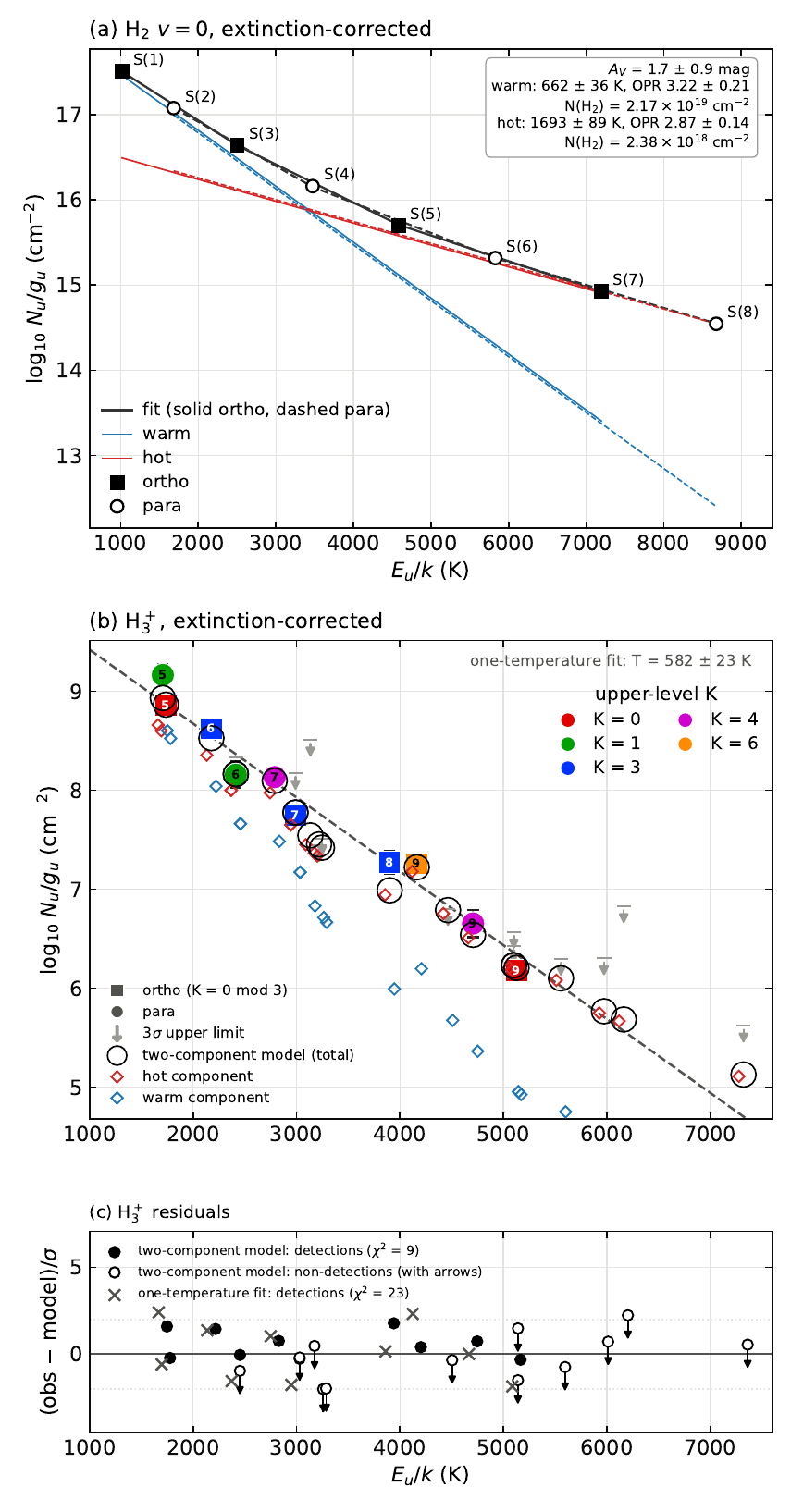}
\caption{(a) H$_2$ rotation diagram (extinction-corrected), with the two-temperature LTE fit and its components. (b) Extinction-corrected
H$_3^+$ rotation diagram of the 22 unblended lines of Table~\ref{tab:fluxes} (ten detections and twelve $3\,\sigma$ upper limits), in the format of \citet{Neufeld2026a}. Colors denote the upper-level
$K$; squares are ortho and circles para levels; the upper-level $J$ is written inside each symbol. Error bars ($1\,\sigma$) are smaller than the symbols where not visible. Dashed line: one-temperature fit to the
ten detections; arrows: $3\,\sigma$ upper limits on the other twelve lines; open circles: the two-component statistical-equilibrium
model of Section~\ref{sec:h2} for $n_{\rm H} = 10^6$\,cm$^{-3}$ and $x_e = 2\times10^{-4}$, with the contributions of its hot (red) and warm (blue) components as small diamonds. (c) Residuals, (observed $-$ model)/$\sigma$, for the two-component model (filled circles: detections; open circles with arrows: non-detections) and for the one-temperature fit to the detections (crosses).\label{fig:excitation}}
\end{figure*}

The H$_2$ rotation diagram (Figure~\ref{fig:excitation}a) is fitted by two components in local thermodynamic equilibrium
(LTE; seven parameters for eight lines): a warm component with
$T = 662 \pm 36$\,K, ortho-to-para ratio (OPR) $3.22 \pm 0.21$ and $N({\rm H_2}) = 2.2\times10^{19}$\,cm$^{-2}$, and a hot component
with $T = 1693 \pm 89$\,K, OPR $2.87 \pm 0.14$ and $N({\rm H_2}) = 2.4\times10^{18}$\,cm$^{-2}$, for $A_V = 1.7 \pm 0.9$\,mag.
This extinction applies to our line of sight, which may pass through the globules themselves; it is unrelated to the much smaller
column between the emitting gas and the ionization front, measured toward the star (Section~\ref{sec:star}).

A one-temperature fit to the H$_3^+$ rotation diagram of the ten unblended detections (Figure~\ref{fig:excitation}b), weighted by
$({\rm S/N})^2$, gives a rotational temperature $T_{\rm rot} = 582 \pm 23$\,K, where the uncertainty has been scaled by
$(\chi^2_r)^{1/2} = 1.7$ for the excess scatter. That scatter shows that no single Boltzmann distribution describes the
populations, which are subthermal (see below), so $T_{\rm rot}$ is a descriptive slope rather than a kinetic temperature. This is similar to the temperature of the warm H$_2$ and
somewhat lower than in HH 168 \citep[730\,K;][]{Neufeld2026a}. The column density in the upper level of the 16.325\um\
line is $N_u = 3.2\times10^{10}$\,cm$^{-2}$ (corrected for extinction) averaged over the aperture, and $(5$--$6)\times10^{10}$\,cm$^{-2}$
in the brightest 3.1\,arcsec$^2$ region of the map. The latter is 2--2.5 times smaller than the value in an aperture of the same size in HH 168 ($1.3\times10^{11}$\,cm$^{-2}$; \citealt{Neufeld2026a}).

The populations are not in LTE. The critical densities $A_{\rm tot}/k_L$ of the upper levels, where $A_{\rm tot}$ is the total
radiative decay rate of the level and $k_L$ the collisional rate coefficient defined below, range from $2\times10^6$ to
$8\times10^7$\,cm$^{-3}$, above the densities of H nuclei, $n_{\rm H} = 10^5$--$10^6$\,cm$^{-3}$, estimated for the globules
\citep{Wesson2024}.

We therefore modeled the excitation with the statistical-equilibrium model of \citet{Neufeld2026a}. It includes the 220 $v = 0$ levels
of H$_3^+$ within 13,700\,cm$^{-1}$ (19,700\,K) of the ground level ($J \le 20$) and all 1219 radiative transitions among them
\citep{Mizus2017}. Collisions are assumed to occur at the Langevin rate, $k_L = 2\times10^{-9}$\,cm$^3$\,s$^{-1}$ for H$_2$ (1.16 and 0.42 times this
for H and He; \citealt{Neufeld2026a}), in the strong-collision limit, in which each collision leaves the ion in a level drawn from a
thermal distribution at the gas temperature. Collision models that favor small changes in $J$ or in energy give
1.4--2.5 times higher ionization rates for the same $x_e$, which we regard as a systematic uncertainty of the excitation model.

We added formation, with a thermal distribution at 3000\,K (the lower end of the 3000--$10^4$\,K range considered by \citealt{Neufeld2026a}, representing ions formed hot by the exothermic reaction
H$_2^+$ + H$_2$ $\to$ H$_3^+$ + H), and destruction at a rate $D = k_{\rm DR}\,n_e + k_{\rm O}\,n({\rm O})$, for dissociative recombination
\citep{McCall2004} and reaction with atomic oxygen \citep{Hillenbrand2022}, with $n_e = x_e\,n_{\rm H}$ and
$n({\rm O}) = 3.2\times10^{-4}\,n_{\rm H}$ \citep{vanHoof2010}. For $x_e = 2\times10^{-4}$, each ion undergoes 140--220
collisions before it is destroyed; this number is independent of $n_{\rm H}$, and differs between the components only through the
temperature dependence of the recombination rate. Formation at 1000 or $10^4$\,K instead of 3000\,K therefore changes $\chi^2$ by $< 0.1$ and the
ionization rates of Section~\ref{sec:zeta} by at most 5\%.

We fitted the 22 unblended lines, detections and non-detections, with two components at the temperatures of the hot and warm H$_2$,
sharing $n_{\rm H}$ and the electron fraction $x_e$, with the H$_3^+$ formation rate of each component free (Figure~\ref{fig:excitation}b). We adopt $x_e = 2\times10^{-4}$,
the electron fraction if the gas-phase carbon (C/H $= 1.6\times10^{-4}$; \citealt{vanHoof2010}) and the low-ionization metals are singly
ionized, as they are at the H/H$_2$ transition of the static models of Section~\ref{sec:cloudy}.
For $n_{\rm H} = 10^6$\,cm$^{-3}$ and $x_e = 2\times10^{-4}$, $\chi^2 = 29$ for 20 degrees of freedom. The ten detected lines contribute 9 to this $\chi^2$; the one-temperature fit to the same ten lines gives $\chi^2 = 23$ (Figure~\ref{fig:excitation}c). The residual scatter exceeds the measurement errors, but the model reproduces the broad
pattern: all twelve non-detections lie within $2.2\,\sigma$ of it, and the hot component supplies about half of the 16.325\um\ line.

\section{The ionization of the \texorpdfstring{H$_2$}{H2}} \label{sec:model}

\subsection{The required ionization rate} \label{sec:zeta}

In steady state, H$_3^+$ is formed at the rate $\zeta\, n({\rm H_2})\, f_3$, where $\zeta$ is the H$_2$ ionization rate and $f_3$
($\approx 1$ in molecular gas) is the fraction of H$_2^+$ ions that react with H$_2$ rather than with H or electrons. If each H$_3^+$
ion emits on average $\eta$ photons in a given line during its lifetime, then for uniform conditions
\begin{equation}
4\pi I/h\nu = \eta\, f_3\, \zeta\, N({\rm H_2});
\end{equation}
in general it is $N({\rm H_2})\langle \eta f_3 \zeta \rangle$, with the product averaged over the H$_2$ column, here
$\sum_i \eta_i f_{3,i} \zeta_i N_i({\rm H_2})$ over the hot and warm components. Since the H$_3^+$ and H$_2$ are measured in the same aperture, the geometry cancels. For the 16.325\um\ line,
$4\pi I/h\nu = 9.7\times10^{7}$\,cm$^{-2}$\,s$^{-1}$ after extinction correction.

The fit of Section~\ref{sec:h2} gives the H$_3^+$ formation rate per unit area of each component, and hence $\zeta$, with $f_3 = 1$
and the observed H$_2$ columns. For $x_e = 2\times10^{-4}$, $\zeta \approx
3\times10^{-12}$\,s$^{-1}$ in the hot H$_2$ ($\eta \approx 7$--10 for the 16.325\um\ line) and $\approx 8\times10^{-13}$\,s$^{-1}$ in
the warm H$_2$ for $n_{\rm H} = 10^6$\,cm$^{-3}$ ($2.5\times10^{-13}$\,s$^{-1}$ for $10^5$\,cm$^{-3}$).

These rates scale with $x_e$, which sets the H$_3^+$ lifetime. In the H$_3^+$-emitting gas, H$^+$ is removed by charge transfer to O,
followed by O$^+$ + H$_2$ $\to$ OH$^+$ + H and further reactions with H$_2$ that end in the dissociative recombination of H$_3$O$^+$
\citep{Stancil1999,Kovalenko2018,Tran2018,Jensen2000}. The electrons therefore come mainly from carbon and the low-ionization metals, whose ionization
state in the warm molecular gas is uncertain. In the hot H$_2$, however, C$^+$ + H$_2$ $\to$ CH$^+$ + H \citep{Hierl1997} probably keeps most of the carbon
neutral; $x_e$ may then be only $(2$--$7)\times10^{-5}$, and with destruction of H$_3^+$ by C and O, $\zeta \approx$
$(0.6$--$1.4)\times10^{-12}$\,s$^{-1}$, with a floor of $2.7\times10^{-13}$\,s$^{-1}$ as $x_e \to 0$. For
$x_e = 2\times10^{-4}$, the hot-layer rate is $\sim 2\times10^4$ times the mean cosmic-ray ionization rate of H$_2$ in diffuse clouds,
$\zeta_t({\rm H_2}) = 1.8\times10^{-16}$\,s$^{-1}$, as revised using C$_2$-based densities \citep{Neufeld2024}.
\rd{In a photoevaporation flow, however, $x_e$ may reach 0.03--0.1 where the H$_3^+$ forms \citep{Henney2007}. Each ion then
undergoes $\lesssim 1.5$ collisions before recombining, and the line pattern is reproduced only if the ions form at $\lesssim 1000$\,K; the
required $\zeta$ is then $5\times10^{-10}$--$3\times10^{-9}$\,s$^{-1}$ (with $f_3 \approx 0.5$--0.75).}

\subsection{Photoionization by the central star} \label{sec:star}

The central star has $T_{\rm eff} = 135$\,kK and $L = 310\,L_\odot$ \citep{Sahai2025,Wesson2024}. We adopted the spectrum of a non-LTE model
atmosphere for these parameters and solar abundances, computed with the T\"ubingen Model-Atmosphere Package \citep[TMAP;][]{Werner2003,Rauch2003,RauchDeetjen2003}
and obtained from the TheoSSA service of the German Astrophysical Virtual Observatory (GAVO) (interpolated between 130 and 140\,kK at $\log g = 8$; \citealt{Sahai2025} likewise adopted solar abundances and $\log g = 8$ in their fit of the stellar spectral energy
distribution).

We computed the H$_2$ ionization rate as a
function of the column density \rd{of H nuclei,} $N_{\rm H}$\rd{,} into the globule, including direct photoionization and secondary ionization by the photoelectrons
\citep{Dalgarno1999}, with the photoionization cross sections of \citet{Verner1996} and \citet{Verner1995}, $\sigma({\rm H_2}) = 2.85\,
\sigma({\rm H})$, the high-energy ratio of \citet{Yan1998} as adopted by \citet{Wilms2000}, and the nebular abundances of \citet{vanHoof2010}, with solar Mg, Si and Fe \citep{Asplund2009}.

\begin{figure*}[!htbp]
\centering\includegraphics[width=0.70\textwidth]{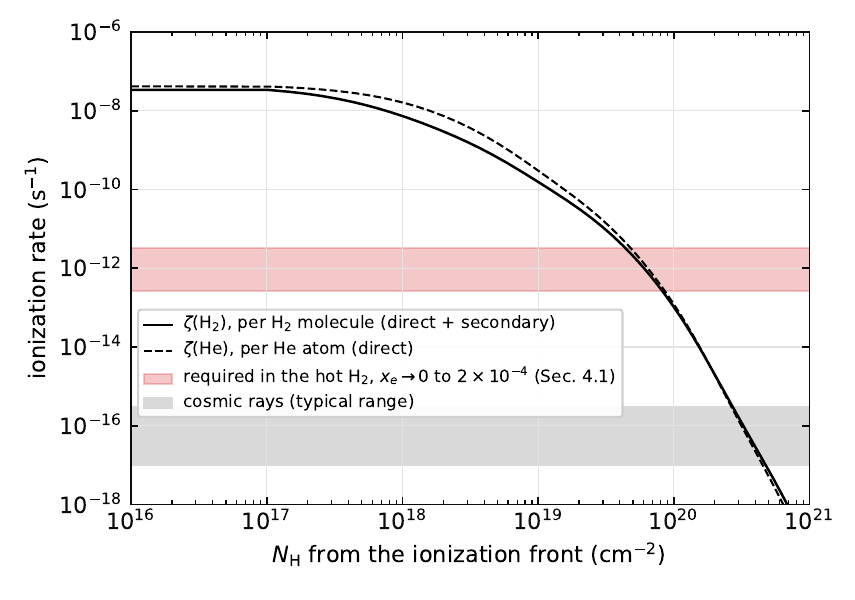}
\caption{Ionization rates due to the central star at the projected distance of the field, as a function of the column density \rd{of H nuclei, $N_{\rm H}$,} from the
ionization front: per H$_2$ molecule (solid, direct and secondary) and per He atom (dashed, direct). Red band: required in the hot H$_2$ \rd{for $x_e \le 2\times10^{-4}$} (Section~\ref{sec:zeta}); gray band: typical cosmic-ray
ionization rates of H$_2$.\label{fig:model}}
\end{figure*}

At the projected distance of the field, $\zeta$ is $1.6\times10^{-10}$\,s$^{-1}$ at $N_{\rm H} = 10^{19}$\,cm$^{-2}$,
$1\times10^{-11}$\,s$^{-1}$ at $3\times10^{19}$\,cm$^{-2}$, $2\times10^{-12}$\,s$^{-1}$ at $5\times10^{19}$\,cm$^{-2}$ and
$1\times10^{-13}$\,s$^{-1}$ at $10^{20}$\,cm$^{-2}$ (Figure~\ref{fig:model}). At $N_{\rm H} = 3$--$5\times10^{19}$\,cm$^{-2}$ photons of
54--100\,eV provide 50--75\% of the ionization and those of 100--300\,eV the rest; at $10^{20}$\,cm$^{-2}$ the 100--300\,eV photons
dominate. A static Cloudy model (Section~\ref{sec:cloudy}) \rd{at constant pressure, with a surface density of $10^4$\,cm$^{-3}$,} gives H$_2$ ionization rates within a factor of 2 of
these values over this range.

The stellar radiation therefore provides the rate required in the hot H$_2$, $3\times10^{-12}$\,s$^{-1}$, at
$N_{\rm H} \approx 4.4\times10^{19}$\,cm$^{-2}$ from the ionization front ($A_V \approx 0.015$\,mag for the dust-to-gas ratio of
\citealt{vanHoof2010}), and that in the warm H$_2$ at $\approx 6\times10^{19}$\,cm$^{-2}$. For the lower electron fractions of Section~\ref{sec:zeta}, the rate required in the hot H$_2$ falls toward the floor of $2.7\times10^{-13}$\,s$^{-1}$, reached at $8\times10^{19}$\,cm$^{-2}$; the hot H$_2$ must therefore lie within \rd{$N_{\rm H} =$} $(4$--$8)\times10^{19}$\,cm$^{-2}$ of the ionization front (Figure~\ref{fig:model}).
\rd{For the high $x_e$ of a photoevaporation flow, the required rate is reached at $N_{\rm H} \approx (2$--$6)\times10^{18}$\,cm$^{-2}$.} These depths refer to the projected
distance. From the stellar He$^+$-ionizing photon rate and the density of the nebular model of \citet{vanHoof2010}, the He$^{++}$
Str\"omgren radius is 0.09\,pc, just inside it: the intervening He$^+$ lowers $\zeta$ by $\sim 15\%$ at these depths, but a globule 25\%
farther from the star would see $\zeta$ lower by a factor of $\sim 5$.

\subsection{Static models} \label{sec:cloudy}

We computed static photoionization models of the star-facing surface of a globule with Cloudy \citep{Chatzikos2023,Gunasekera2023},
for the stellar spectrum of Section~\ref{sec:star} at the projected distance\rd{, at constant gas pressure, as expected for globules small
enough to be in pressure equilibrium. Surface densities of $10^3$, $3\times10^3$ and $10^4$\,cm$^{-3}$ give
$n_{\rm H} = (2$--$7)\times10^5$\,cm$^{-3}$ at the H/H$_2$ transition, as estimated for the globules} \citep{Wesson2024}. The models include Cloudy's full chemical network and its detailed model of the H$_2$ molecule
\citep{Shaw2005}, the nebular abundances and dust-to-gas mass ratio (0.0062) of \citet{vanHoof2010}, with Cloudy's planetary-nebula
abundances for the other elements and its interstellar grain size distribution (without PAHs; in the Ring Nebula the PAH emission is weak and arises mainly outside the dense clumps, \citealt{Clark2025}), a cosmic-ray ionization rate
$\zeta({\rm H_2}) \approx 10^{-16}$\,s$^{-1}$ and a microturbulent velocity of 3\,\kms, and extend to $N_{\rm H} = 2\times10^{21}$\,cm$^{-2}$.

The temperature is computed from thermal balance, and we computed the H$_3^+$ lines from the H$_3^+$ formation and destruction rates of
each model with the statistical-equilibrium model of Section~\ref{sec:h2}, as intensities normal to the front. The H/H$_2$ transition, where half of the H nuclei are in H$_2$, lies at $N_{\rm H} = \rd{3}$--$\rd{4}\times10^{19}$\,cm$^{-2}$ \rd{from the ionization front}, where the
stellar ionization rate is close to that required, but in gas at \rd{180--370}\,K. The models underpredict the H$_2$ S(1) line by factors of
\rd{55--250} and the 16.325\um\ line by factors of \rd{120--370}. The H$_3^+$ forms at nearly the required rate (\rd{$2.4\times10^7$}\,cm$^{-2}$\,s$^{-1}$
for \rd{a surface density of $10^4$}\,cm$^{-3}$, against $2.5\times10^7$\,cm$^{-2}$\,s$^{-1}$ in the fit of Section~\ref{sec:h2}), with
$x_e \approx \rd{2}$--$\rd{3}\times10^{-4}$ and most of the carbon ionized \rd{at the transition}, but in the cold gas each ion emits little more than its formation cascade (so the predicted line scales with the
assumed formation temperature; a lower one would make it weaker still). \rd{Constant-density models ($n_{\rm H} = 10^5$--$10^6$\,cm$^{-3}$)
give similar results, with the S(1) and 16.325\um\ lines underpredicted by factors of 20--90 and 60--230.}

Stellar heating thus does not maintain molecular gas at the observed
660--1700\,K in these models: the H$_2$ column above 500\,K is $\lesssim 10^{17}$\,cm$^{-2}$, less than 1\% of the observed warm H$_2$.

The warm H$_2$ and its H$_3^+$ thus suggest processes that static models omit. The most plausible is advection. In a merged
ionization and dissociation front \citep{Bertoldi1996}, the photoevaporation flow carries H$_2$ toward the ionization front faster
than it is destroyed; in advective models of the Helix Nebula knots, the H$_2$ then survives into gas heated to 1000--2000\,K by the
stellar EUV radiation \citep{Henney2007}. This would place warm H$_2$ where the stellar ionization rate is high: within the depth found in
Section~\ref{sec:star}, where in the static models the H$_2$ is either absent or cold.

Warm H$_2$ in PNe has also often been attributed to shocks \citep{Zuckerman1988}. In HH 168, the best-fitting shock model, a 60\,\kms\ dissociative shock, forms H$_3^+$ through H$^+$ + H$_2$ $\to$ H$_2^+$ + H
\citep{Neufeld2026a}, a reaction endothermic by 1.8\,eV. A photoevaporating globule is compressed by a much slower shock that runs ahead of its ionization front
\citep{Bertoldi1989}: for pressure balance with the ionized gas at the front, of density $\lesssim F_0/c_{\rm II} \approx
1.6\times10^4$\,cm$^{-3}$ for the stellar ionizing photon flux $F_0 = 1.8\times10^{10}$\,cm$^{-2}$\,s$^{-1}$ and a sound speed
$c_{\rm II} \approx 11$\,\kms, a globule with $n_{\rm H} \geq 10^5$\,cm$^{-3}$ is entered by a shock of $\lesssim 4$--7\,\kms. That is too
slow to ionize the gas. The
16.325\um/S(6) flux ratio here is also $\sim 3$ times that in HH 168. The ionization is therefore most plausibly stellar.

\section{Discussion} \label{sec:disc}

The H$_3^+$ rotational lines in the Ring Nebula thus arise in the warm and hot molecular skin of the dense globules, where H$_2$ is ionized
by the central star at rates up to $\sim 2\times10^4$ times those due to cosmic rays. The detection extends the H$_3^+$ rotational spectrum from
shocked gas \citep{Neufeld2026a} to photoionized molecular gas, and makes H$_3^+$ a probe of the extreme-ultraviolet and soft X-ray
irradiation of molecular gas by \rd{very} hot stars, as \citet{Draine1990} anticipated for X-ray-irradiated gas in starburst galaxies.

The same irradiation forms the OH$^+$ detected in the Ring Nebula with Herschel, mainly through O$^+$ + H$_2$
(Section~\ref{sec:zeta}; \citealt{Aleman2014}), consistent with its presence only in PNe with very hot central stars.

The H$_3^+$ emission follows the low-ionization lines ([Ne\,{\sc ii}], [Ar\,{\sc ii}] and [Fe\,{\sc ii}]) and avoids the high-ionization lines ([Ne\,{\sc v}] and [S\,{\sc iv}]; Section~\ref{sec:lines}), as expected if it arises just behind the ionization fronts of the globules that face the star. A detailed analysis of the ionic emission is deferred to future work.

How the molecular gas is heated to the observed temperatures, how close to the ionization front it lies, and how much of the
carbon in it is ionized, are the main open questions. Dynamical models of photoevaporating globules that follow the heating, the chemistry and the H$_3^+$ excitation together
will be needed to answer them. Spatially resolved observations of the globule surfaces, and deeper NIRSpec spectra of the $\nu_2$ band,
which is not detected in the existing 584\,s NIRSpec G395M observation of program GO 1558, would test the location of the H$_3^+$ relative to the ionization front.

\section{Summary} \label{sec:summary}

\begin{enumerate}\setlength{\itemsep}{0pt}\setlength{\parskip}{0pt}\setlength{\parsep}{0pt}
\item We have detected the forbidden pure rotational spectrum of H$_3^+$ in archival JWST/MIRI observations of the Ring Nebula, to our
knowledge the first detection of H$_3^+$ in a planetary nebula. The 16.325\um\ line is detected at $32\,\sigma$, along with nine further
unblended lines.
\item The H$_3^+$ emission spatially follows the H$_2$ emission of the dense globules and lies at the H$_2$ velocity to within $\sim 7$\,\kms. The
H$_2$ rotational lines show warm (660\,K) and hot (1700\,K) components. A statistical-equilibrium model at these temperatures reproduces the
subthermal H$_3^+$ line pattern, with much of the H$_3^+$ emission arising in the hotter (1700\,K) H$_2$ component.
\item The H$_3^+$ emission requires an H$_2$ ionization rate of $\sim 3\times10^{-12}$\,s$^{-1}$ in the hot H$_2$
for an electron fraction of $2\times10^{-4}$. The rate
is $\sim 2\times10^4$ times the mean cosmic-ray ionization rate in diffuse clouds.
\item The extreme-ultraviolet and soft X-ray radiation of the central star provides this rate within \rd{an H-nucleus column density of} $(4$--$8)\times10^{19}$\,cm$^{-2}$ of the
ionization front of the globules\rd{, or the higher rate required for the high electron fractions of a photoevaporation flow within
$(2$--$6)\times10^{18}$\,cm$^{-2}$}.
\item Static photoionization models place the H$_2$ in gas too cold to excite the H$_3^+$ lines and do not reach the observed H$_2$
temperatures. The most plausible explanation is the advection of H$_2$ through a photoevaporation front into gas heated
by the stellar EUV radiation; the shocks driven into the globules are too slow to ionize the gas.
\item The H$_3^+$ rotational lines thus provide a new probe of the photoionization of molecular gas by \rd{very} hot stars.
\end{enumerate}

\begin{acknowledgments}
This work is based on observations made with the NASA/ESA/CSA James Webb Space Telescope, obtained from the Mikulski Archive for Space
Telescopes at the Space Telescope Science Institute, which is operated by the Association of Universities for Research in Astronomy,
Inc., under NASA contract NAS 5-03127 for JWST. These observations are associated with program 1558. \rd{The specific observations analyzed can be accessed via
\dataset[doi:10.17909/xv1f-fe02]{https://doi.org/10.17909/xv1f-fe02}.} We thank the GO 1558 team for designing these
observations\rd{, and the referee for useful comments that improved the paper}. D.A.N.\ was supported by NASA through grants JWST-AR-06829 (JWST Theory program, ``Understanding the rich molecular spectra
observed with NIRSpec and MIRI/MRS toward protostellar outflows'') and JWST-GO-07084 from the Space Telescope Science Institute,
which is operated by the Association of Universities for Research in Astronomy, Inc., under NASA contract NAS 5-03127. The TheoSSA service (\url{http://dc.g-vo.org/theossa}) used to retrieve
theoretical spectra for this paper was constructed as part of the activities of the German Astrophysical Virtual Observatory.

Use of generative AI: Anthropic's Claude (Opus 5.5, accessed through Claude Code) was used for data analysis, coding,
running and analyzing the models, generating the figures from the data using Matplotlib, and drafting and revising the manuscript text.
OpenAI's Codex was used to write the initial line-fitting and mapping code. The authors take full responsibility for the content of this
paper.
\end{acknowledgments}

\facility{JWST (MIRI, NIRSpec)}
\software{JWST Calibration Pipeline (version 2.0.1; \citealt{Bushouse2026}), Cloudy (version C23.01; \citealt{Chatzikos2023,Gunasekera2023}), Astropy \citep{Astropy2022}, NumPy \citep{Harris2020}, SciPy \citep{Virtanen2020}, Matplotlib \citep{Hunter2007}}

\clearpage
\appendix
\section{Line fluxes} \label{app:tables}

\startlongtable
\begin{deluxetable*}{lcccc}
\tabletypesize{\footnotesize}
\tablecaption{Line fluxes in the 15.7\,arcsec$^2$ aperture common to all MIRI/MRS sub-bands (MRS position 1)\label{tab:fluxes}}
\tablehead{\colhead{Line} & \colhead{$\lambda$} & \colhead{$E_u/k$} & \colhead{Flux} & \colhead{$\sigma$} \\
\colhead{} & \colhead{($\mu$m)} & \colhead{(K)} & \colhead{($10^{-17}$\,erg\,cm$^{-2}$\,s$^{-1}$)} & \colhead{($10^{-17}$\,erg\,cm$^{-2}$\,s$^{-1}$)}}
\startdata
H$_2$ 0--0 S(1) & 17.0348 & 1015 & 963.8 & 5.64 \\
H$_2$ 0--0 S(2) & 12.2786 & 1682 & 1175.5 & 8.78 \\
H$_2$ 0--0 S(3) & 9.6649 & 2504 & 6577.9 & 45.61 \\
H$_2$ 0--0 S(4) & 8.0250 & 3474 & 3203.0 & 39.94 \\
H$_2$ 0--0 S(5) & 6.9095 & 4586 & 10085.3 & 228.63 \\
H$_2$ 0--0 S(6) & 6.1086 & 5829 & 3443.1 & 39.48 \\
H$_2$ 0--0 S(7) & 5.5112 & 7197 & 9294.3 & 78.47 \\
H$_2$ 0--0 S(8) & 5.0531 & 8677 & 2494.7 & 29.72 \\
\hline
H$_3^+$ $(7,2)\to(6,1)$$^*$ & 19.9567 & 3134 & 6.86 & 8.62 \\
H$_3^+$ $(6,1)\to(5,2)$$^*$ & 18.0569 & 2413 & 2.97 & 2.77 \\
H$_3^+$ $(7,3)\to(7,6)$ & 18.0018 & 2990 & 9.30 & 2.26 \\
H$_3^+$ $(6,3)\to(6,6)$ & 17.1985 & 2177 & 19.21 & 2.66 \\
H$_3^+$ $(5,0)\to(4,3)$ & 16.3254 & 1737 & 30.00 & 0.93 \\
H$_3^+$ $(7,1)\to(6,2)$$^*$ & 16.1011 & 3218 & 1.34 & 1.29 \\
H$_3^+$ $(9,4)\to(9,7)$ & 15.7859 & 4707 & 2.43 & 0.78 \\
H$_3^+$ $(7,4)\to(7,7)$ & 14.2794 & 2789 & 9.85 & 1.06 \\
H$_3^+$ $(10,5)\to(10,8)$$^*$ & 13.7123 & 5557 & 0.99 & 0.85 \\
H$_3^+$ $(9,1)\to(8,2)$$^*$ & 13.4998 & 5101 & 0.38 & 0.90 \\
H$_3^+$ $(7,0)\to(6,3)$$^*$ & 13.4582 & 3246 & 3.52 & 7.80 \\
H$_3^+$ $(5,1)\to(4,4)$ & 13.3642 & 1707 & 9.00 & 2.35 \\
H$_3^+$ $(9,5)\to(9,8)$$^*$ & 12.9717 & 4465 & 3.31 & 1.28 \\
H$_3^+$ $(10,1)\to(9,2)$$^*$ & 12.6096 & 6164 & 12.24 & 5.00 \\
H$_3^+$ $(6,1)\to(5,4)$ & 12.3163 & 2413 & 6.17 & 1.90 \\
H$_3^+$ $(9,0)\to(8,3)$ & 11.7384 & 5127 & 7.33 & 1.48 \\
H$_3^+$ $(9,6)\to(9,9)$ & 10.7912 & 4163 & 14.65 & 2.89 \\
H$_3^+$ $(11,0)\to(10,3)$$^*$ & 10.6620 & 7322 & 4.34 & 2.88 \\
H$_3^+$ $(9,1)\to(8,4)$$^*$ & 10.3040 & 5101 & 4.56 & 1.59 \\
H$_3^+$ $(7,3)\to(6,6)$$^*$ & 8.7244 & 2990 & 5.47 & 5.49 \\
H$_3^+$ $(8,3)\to(7,6)$ & 8.4099 & 3901 & 12.17 & 3.30 \\
H$_3^+$ $(10,3)\to(9,6)$$^*$ & 7.9497 & 5973 & 6.23 & 3.91 \\
\enddata
\tablecomments{Observed fluxes, not corrected for extinction. H$_2$: Gaussian fits of free center and width. H$_3^+$: fixed center
(velocity of the 16.325\um\ line) and fixed width (Section~\ref{sec:obs}), for the 22 unblended lines of Table 1 of
\citet{Neufeld2026a}; uncertainties include the empirical calibration of Section~\ref{sec:obs}. $^*$Not detected above $3\,\sigma$.
H$_3^+$ wavelengths and $E_u/k$ are from MiZATeP \rd{\citep{Mizus2017,Bowesman2023}}. The 10.791 and 14.279\um\ lines were fitted simultaneously with H\,{\sc i} 16--9 and with an unidentified line at 14.296\um, respectively (Figure~\ref{fig:spectra}).}
\end{deluxetable*}

\bibliography{refs}{}
\bibliographystyle{aasjournal}

\end{document}